# Transition from Optically Excited to Intrinsic Spin Polarization in WSe$_2$

S. Hedwig[1], G. Zinke[1], J. Braun[2], B. Arnoldi[1], A. Pulkkinen[3], J. Minár[3], H. Ebert[2], M. Aeschlimann[1,a], B. Stadtmüller[4,b]

[1]Department of Physics and Research Center OPTIMAS, Rheinland-Pfälzische Technische Universität Kaiserslautern-Landau, 67663 Kaiserslautern, Germany

[2]Department Chemie, Physical Chemistry, Ludwig-Maximilians-Universität München, 81377 München, Germany

[3]New Technologies Research Centre, University of West Bohemia, 301 00 Plzeň, Czech Republic

[4]Institute of Physics, Experimentalphysik II, Augsburg University, 86159 Augsburg, Germany

[a]Contact author: ma@physik.uni-kl.de
[b]Contact author: benjamin.stadtmueller@uni-a.de

**Abstract:**

Layered 2D van der Waals materials, such as transition metal dichalcogenides, are promising for nanoscale spintronic and optoelectronic applications. Harnessing their full potential requires understanding both intrinsic transport and the dynamics of optically excited spin and charge carriers – particularly the transition between excited spin polarization and the conduction band's intrinsic spin texture. Here, we investigate the spin polarization of the conduction bands of bulk WSe$_2$ using static and time-resolved spin-resolved photoemission spectroscopy, complemented by photocurrent calculations. Electron doping reveals the intrinsic spin polarization, while time-resolved measurements trace the evolution of excited spin carriers. We find that intervalley scattering is spin-conserving, with spin transport initially governed by photoexcited carriers and aligning with the intrinsic conduction band spin polarization after ~150 fs.

Transition metal dichalcogenides (TMDs) have emerged as a prominent class of two-dimensional materials with remarkable electronic, optical, and spintronic properties, making them attractive for both fundamental research and technological applications [1–4]. Among these materials, $WSe_2$ has garnered significant attention due to its unique spin-related phenomena, including large spin-orbit coupling (SOC) and valley-dependent spin properties, which arise from its broken inversion symmetry and strong SOC in the transition metal atoms. As a bulk crystal, $WSe_2$ exhibits an indirect bandgap between the valence band maximum at K/K′ and the Σ valley of the conduction band, whereas the monolayer features a direct bandgap at the K and K′ points [5]. Importantly, the strong SOC in both monolayer and bulk $WSe_2$ induces substantial spin splitting, particularly in the valence band, creating spin polarization locked to specific valleys (K and K′) – a phenomenon known as spin-valley coupling [6, 7]. Combined with local symmetry breaking, this leads to highly spin-polarized bands that globally compensate due to inversion symmetry, resulting in a Kramers degenerate spin polarization across the Brillouin zone – an effect called "hidden spin polarization" [8, 9]. This enables spin functionalities without magnetization, offering potential for future low-power spin-based technologies [10]. While the valence band spin properties of bulk $WSe_2$ have been extensively studied [9, 11–13], the spin texture of the conduction band remains largely unexplored [14]. Only recently was the spin polarization of excitons directly measured in a photoemission experiment [15]. However, excitonic spin polarization does not necessarily reflect the intrinsic spin polarization of the conduction band, which governs electronic charge transport in such materials. Optically excited populations undergo various scattering processes leading to energy and spin relaxation and redistribution in momentum space, thus defining carrier transport on a microscopic level. An important question in the framework of optoelectronic and spintronic applications is how ultrafast spin polarization of excited carriers relates to the intrinsic spin polarization of the transport states. Equally important is the transition between these regimes, reflecting the transition from ballistic to diffusive transport. A previous study tracked the spin polarization decay of photoexcited bright excitons at the K and K′ points using time- and spin-resolved ARPES [15]. Interestingly, they observed a time-independent spin-down polarization of intervalley scattered excitons at the Σ valley, independent of the excitation light helicity, highlighting their potential as local spin reservoirs for spin injection in TMD heterostructures. However, they did not observe any temporal transition at the Σ valleys from the initially excited to the steady scattered ensemble spin polarization.

In this study, we aim to shed new light on the spin polarization of the conduction band of bulk WSe$_2$. In particular, we compare the intrinsic single-particle spin polarization of the conduction band and the optically excited spin carrier dynamics and their respective relaxation time scales. Recent advancements in experimental techniques, such as time- and spin-resolved pump-probe photoemission spectroscopy, offer a powerful approach to directly probe the spin texture and dynamics in the conduction band. These techniques enable time-resolved mapping of spin polarization in momentum space, allowing deeper insights into ultrafast spin relaxation mechanisms and spin-dependent scattering processes in WSe$_2$. In combination with static spin-resolved measurements performed on alkali-metal doped WSe$_2$ and photoemission calculations, this allows us to directly access and disentangle the intrinsic and optically excited spin polarization of conduction band electrons.

We start with the single-particle spin polarization of the conduction band, defining the diffusive transport properties. We first compare the calculated spin-resolved photoemission intensity and the band structure obtained in static spin-resolved photoemission experiments on an alkali-metal doped WSe$_2$ crystal. The performed photocurrent calculations (see methods section for further details) take into account final state effects, enhancing comparison with the experimental data. Figure 1 depicts the calculated photoemission intensity (a) and the corresponding out-of-plane ($P_z$) component of the calculated spin polarization (b) for a photoemission process with $hv = 21$ eV along the Γ–K direction of a bulk WSe$_2$ crystal. The energy axis is referenced to the maximum of the valence band, which shows a strong spin-orbit splitting of $(0.48 \pm 0.04)$ eV at the K and K′ points around $k_\parallel = \pm (1.28 \pm 0.02)$ Å$^{-1}$, which is, as well as the overall band dispersion, in very good agreement with previous experimental findings and DFT calculations [16–18,9]. As expected from time reversal symmetry, at the K and K′ points the spin polarization is inverted. The local conduction band minima at the K and K′ points also exhibit a considerable spin splitting of $(80 \pm 4)$ meV and follow the spin sequence of the valence bands. At the Σ points at $k_\parallel = \pm (0.69 \pm 0.02)$ Å$^{-1}$, the conduction band shows a global minimum at $E - E_{\text{VBmax}} = (1.05 \pm 0.02)$ eV and a spin polarization opposite to that of the lower conduction band branch at the respective K point. This energy also marks the electronic transport gap of the calculated band structure.

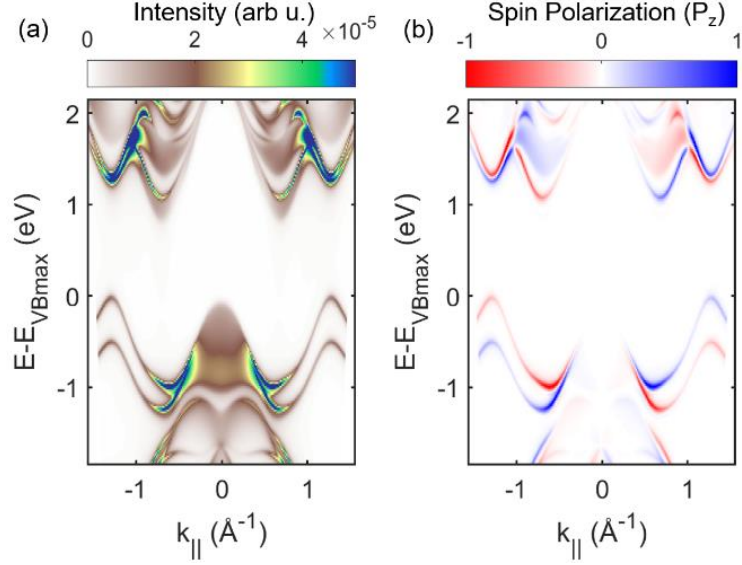

Figure 1: Calculated photoemission spectra of bulk $WSe_2$ along the Γ–K direction. (a) Spin-integrated intensity and (b) out-of-plane spin polarization $P_z$, with blue (red) marking spin-up (-down) spin polarization. The energy axis is referenced to the valence band maximum.

To experimentally access the single-particle spin polarization of the conduction band at the high-symmetry K and Σ points, we performed static spin- resolved ARPES measurements on a potassium-doped $WSe_2$ sample. The doping results in an electron transfer from the alkali metal to the $WSe_2$ semiconductor, raising the chemical potential and ultimately populating the conduction band. A discussion of the changes in the ARPES spectrum upon potassium doping and further characterization of the doped band structure is found in the appendix of this work.

Figures 2 (a) and (b) show the measured spin-resolved photoemission signal at the conduction band onset at the K and K′ points, respectively, with valence band spectra as insets for reference. The spin polarization at the Σ and Σ′ points is shown in Figures 2 (c) and (d). All spectra were obtained for the highest potassium doping concentration as discussed in the appendix. Blue (red) markers represent measured spin-up (-down) signal, and solid lines show the smoothed signal. The different symmetry points are visualized within a sketch of the hexagonal surface Brillouin zone in the corner of each plot. With the used photon energy of 21.2 eV, the photoelectrons have a rather small inelastic mean free path of ~1 nm in $WSe_2$, so that mainly the first two (tri-)layers contribute to the measured spin signal with a ratio of around 3:1 between first and second layer [9, 19, 20].

It can be seen that the spin polarization of the conduction band at the K points (Figures 2 (a) and (b)) follows that of the upper valence bands and is positive at the K and negative at the K′ point. The conduction band at the Σ and Σ′ point (Figure 3 (c) and (d)) also shows an opposing spin polarization, which is negative at the Σ and positive at the Σ′ point, opposite to the spin polarization observed at the corresponding K points. Overall, the spin polarization of the valence and conduction band at the four measured symmetry points shows the alternating behavior in its sign, as expected from time reversal symmetry.

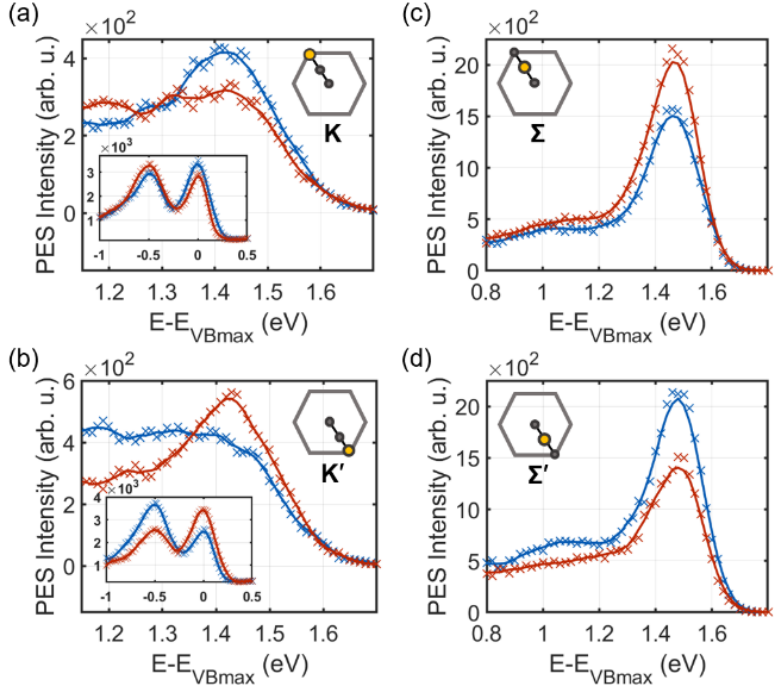

Figure 2: (a)-(b) Spin-resolved spectra of the conduction band of potassium doped $WSe_2$ at the K (a) and K′ point (b) of the Brillouin zone (sketch in the corner of each graph). The corresponding valence band spectra are shown as insets. (c)-(d) Spin-resolved spectra of the conduction band at the Σ (c) and the Σ′ (d) point. Blue (red) markers refer to the measured spin-up (-down) data points, the same-colored solid lines represent the smoothed signal for better visualization.

To compare the findings for the intrinsic spin polarization with those of excited spin carrier populations, we conducted time-, angle-, and spin-resolved pump-probe photoemission experiments. This enables complementary access to the conduction band of $WSe_2$ and allows to probe the ultrafast spin polarization dynamics after optical excitation. In particular, electrons in the valence band are optically excited into previously unoccupied conduction states at the K point, using an ultrashort circularly polarized NIR laser pulse (< 40 fs, 1.59 eV). The evolution of the excited system is then probed by a second time-delayed ultrashort XUV probe pulse (< 30 fs, 22.3 eV [21]), ultimately photoemitting the excited charge carriers. With applied pump fluencies up to 6.2 mJ/cm², lying in an excitation regime well above the critical Mott density where a transition between exciton gas and electron-hole plasma takes place [22, 23], we expect to excite mostly free carriers in the conduction band rather than excitons. By using left (σ⁻) and right (σ⁺) circularly polarized pump pulses, spin-up or spin-down carriers at the K points can be selectively excited into the conduction band [24–26, 7, 14]. The measurements were conducted at identical K and Σ points along the Γ–K direction with varying pump light polarization, as illustrated in Figures 3 (a) and (b). Due to the high surface sensitivity of the XUV probe light, the combined photoemission signal of only the first (L1) and second (L2) layer is obtained, as mentioned before. Along the observed Γ–K direction, this means that with σ⁻-polarized pump light, all K points in the first and second layer are excited but mainly signal from the first layer is obtained, whereas for σ⁺-polarized pulses, all K′ points of the first two layers are addressed but the measured signal stems mostly from the second layer. The investigated points in the surface Brillouin zone are outlined by green areas in Figures 3 (a) and (b), while the light polarization and layer-dependent contribution to the photoemission signal is illustrated by a yellow corona around the excited K/K′ points. Figure 3 (c) depicts the excitation scheme for an exemplary excitation with σ⁻-polarized light. Upon selective optical excitation of spin carriers from the valence band to the conduction band at the K point (1.), a subsequent intervalley scattering process from the K to the Σ point takes place (2.). One important goal here is to elucidate the spin polarization and its evolution, especially after scattering to the Σ point, to gain insight into the relaxation of the non-equilibrium spin population. Known mechanisms that lead to depolarization of the valleys include intervalley exchange coupling between neighboring K and K′ points, whereby both electron and hole are transferred, individually flipping their spin [27–30], as well as phonon-mediated relaxation processes [31–33]. In the following, we discuss the results for the conduction band. Corresponding time- and spin-resolved measurements of the valence band, which reflect the hole dynamics, are presented in Figure S1 in the SI [34]. Figure 3 (d) shows the measured spin signal at the K point of the conduction band in the left and the Σ point in the right column for three time steps (0 fs, 70 fs and 750 fs) after optical excitation with σ⁻-polarized light. Blue (red) dots resemble the measured spin-up (-down) data points, whereas the lines of the same color are corresponding Gaussian fits. At 0 fs, marking the onset of the excitation, a ~75% spin-up polarized population at the K point can be observed, while no significant population has yet scattered to the Σ point. 70 fs later, the population at the K point has exceeded its maximum with still ~70% spin-up polarization. The small amount of spin-down electrons observed in the signal most likely stems from a combination of spin-flip scattered and backscattered carriers due to intervalley exchange coupling between the K and K′ point, as well as an imperfect light polarization, leading to an additional weak excitation of spin-down carriers at the K′ points of the second layer. At the same time (70 fs), a population of the same spin-up polarization is found at the Σ point. This indicates that the initial scattering from the K to the Σ point is spin-conserving. 750 fs after optical excitation, nearly all excited carriers at the K point have either recombined or scattered to the neighboring valley at the Σ point. Remarkably, the population at the Σ point shows a complete reversal of the sign of spin polarization 750 fs after the initial excitation, with a surplus of spin-down carriers.

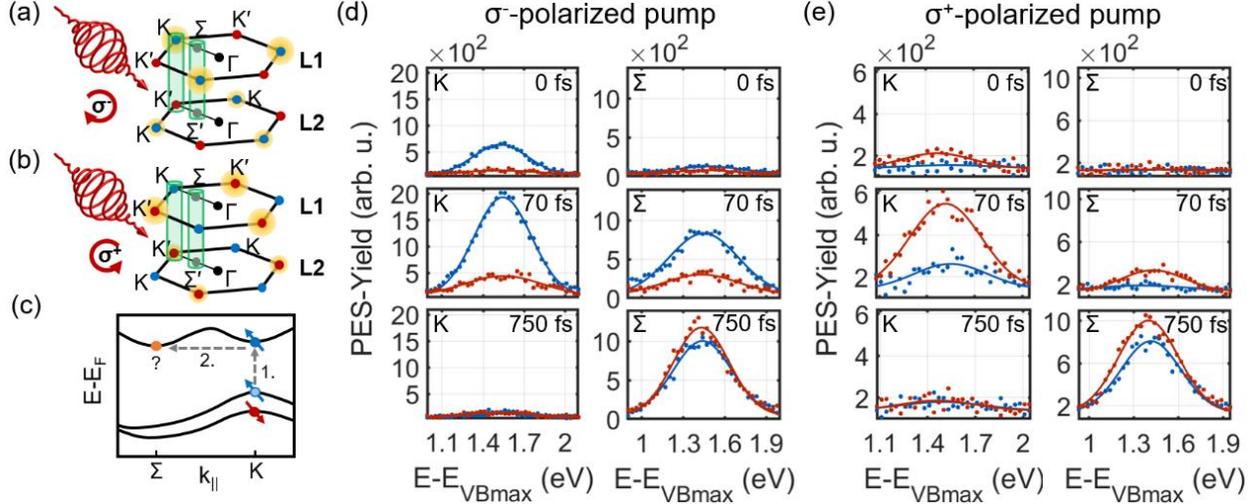

Figure 3: Sketch of the hexagonal surface Brillouin zone of the first (L1) and second (L2) layer of WSe$_2$ for excitation with σ⁻ (a) and σ⁺-polarized (b) pump pulses. Depending on pump helicity, spin carriers are excited at different points, indicated by yellow coronae. Their size reflects the layer-dependent contribution to the signal. Measurements were taken at the same positions marked in green. (c) Excitation scheme: (1) Spin-selective excitation of carriers from the valence to conduction band at the K point; (2) Intervalley scattering to the Σ point. (d)-(e) Spin-up (blue) and spin-down (red) photoemission intensities and Gaussian fits at 0 fs, 70 fs, and 750 fs after σ⁻ (d) and σ⁺ (e) excitation. Left and right columns show signals at the K and Σ points, respectively

The same experiment was repeated with σ⁺-polarized light. In this case, mainly the K′ point of the second layer in the observed region is excited. The results are shown in Figure 3 (e). After a spin-selective excitation of spin-down carriers at the K′ point of the second layer, the initial scattering towards the Σ′ valley again proves to be spin-preserving, confirming the observation made with σ⁻-polarized excitation in Figure 3 (d). 750 fs after optical excitation, no change in the sign of the spin polarization is observed at the Σ′ point. Instead, a finite spin-down polarization is present, similar to that observed for σ⁻-polarized excitation in Figure 3 (d), indicating that this corresponds to the intrinsic spin polarization at the Σ valley. The same spin-down polarization has recently also been reported for scattered dark excitons at the Σ valley following initial excitation at the K point with σ⁻- and σ⁺-polarized light [15]. However, in that study, no thermalization of the scattered spin carriers was observed; instead, a time-independent spin-down polarization persisted. We further conducted additional time- and spin-resolved measurements at the Σ′ point of the first layer after excitation with σ⁻-polarized light (see Figure S2 in the SI [34]). On longer time scales (750 fs) after optical excitation, the scattered population exhibits a clear spin-up polarization. This is in very good agreement with the intrinsic spin polarizations at the Σ and Σ′ valley of the conduction band observed on potassium-doped $WSe_2$, shown in Figures 2 (c) and (d). This consistency between the two complementary approaches allows the spin polarization of the scattered population on longer timescales to be directly related to the intrinsic single-particle spin polarization of the conduction band. The equalization of the spin polarization can further be associated with carrier thermalization at the Σ point, as observed in additional spin-integrated and highly time-resolved measurements (see Figure S3 in the SI [34]), enabling a classification of the relevant time scales.

In conclusion, our time-, angle-, and spin-resolved photoemission experiments provide insight into the nature of the spin polarization at the conduction band valleys of $WSe_2$ as well as the evolution of optically excited non-equilibrium spin populations. After a spin selective initial excitation of spin carriers at the K points, depending on the pump light helicity, the subsequent intervalley scattering to the Σ points appears to be spin-conserving. For later times after the scattering, a finite spin-down (-up) polarization is observed at the Σ (Σ′) valley, which is independent from the initially excited and scattered spin carriers. This hints at a SOC mediated and spin flip involving thermalization of the scattered non-equilibrium populations at the Σ points and an equalization of the spin polarization due to interaction of neighboring Σ points, as suggested in previous work [14]. Comparing these findings with the spin polarization of potassium doped $WSe_2$, it can be seen that in both cases the spin polarization exhibits the same behavior with a finite spin polarization at the respective Σ points of the conduction band. As a result, this indicates an equalization of the optically excited and initially spin-conserved scattered spin carriers at the Σ valleys toward the intrinsic spin polarization of the conduction band. For spin transport in optoelectronic or spintronic applications, this means that on short time scales ($\lesssim 150$ fs) the spin state of excited spin carriers is preserved even after phonon-mediated intervalley scattering. This is particularly important for spin-conserving ballistic transport processes at interfaces, where the optically excited carriers thus define the spin properties of the system, which not necessarily reflect the intrinsic spin polarization of the material they enter. The latter becomes important at longer times ($\gtrsim$ 150 fs), when the excited spin carriers have equilibrated to the intrinsic spin polarization via spin relaxation processes. For diffusive transport involving frequent scattering events, the essential spin characteristics are therefore governed by the intrinsic single-particle spin polarization of the material.

**Methods:**

All photoemission experiments were performed at room temperature under UHV conditions (~$10^{-10}$ mbar) using a hemispherical energy analyzer (SPECS Phoibos 150) in combination with a CCD detector for ARPES and a commercial VLEED spin filter system (Focus FERRUM) [41] for 1D spin-resolved measurements. The spin-resolved data was taken for the out-of-plane spin component. As light sources, we used the He I$\alpha$ line (21.2 eV) of a monochromatic He gas discharge lamp (Scienta VUV5k) for static measurements and a high harmonic generation (HHG) setup providing linearly polarized fs-pulsed XUV radiation (< 30 fs, 22.3 eV) for the time resolved experiments. For the HHG process, the frequency doubled (390 nm) output of a Ti:Sa laser amplifier system (780 nm, 10 kHz, < 40 fs) is used [21]. The fundamental of the amplifier system (1.59 eV) is used for the optical excitation of the sample in a pump-probe interferometric scheme. The spatial overlap of the pump and probe beam was optimized directly on the sample before each experiment and was actively stabilized during the measurements. The pump light helicity can be selected using an additional quarter wavelength plate in the beam path.

The 2H-WSe$_2$ samples were purchased from HQ graphene and were cleaved in situ under UHV condition to obtain a flat and clean surface. Furthermore, potassium doping of the samples was realized under UHV conditions using commercially available alkali-metal dispensers.

In addition to our experimental data, we conducted self-consistent electronic structure calculations within the ab initio framework of density functional theory. For these calculations, we employed the Vosko, Wilk, and Nusair parametrization to account for the exchange and correlation potential [42]. The electronic structure was determined in a fully relativistic mode by solving the Dirac equation, utilizing the relativistic multiple-scattering formalism, also known as the Korringa-Kohn-Rostoker (KKR) method, in the full potential KKR mode [43-45].

Photocurrent calculations were based on the resulting half-space electronic structure, which was represented by single-site scattering matrices for the various layers. These matrices, along with the wave functions corresponding to the initial and final state energies, were used as input quantities.

**Acknowledgements:**

The experimental work was funded by the Deutsche Forschungsgemeinschaft (DFG, German Research Foundation) - TRR 173 - 268565370 Spin+X: spin in its collective environment (Projects A02).

J. M. and A. P. thank the project Quantum materials for applications in sustainable technologies (QM4ST), funded as project No. CZ.02.01.01/00/22_008/0004572 by P JAK, call Excellent Research.

**Data availability:**

The data supporting the findings of this study are available from the corresponding authors upon request

# Appendix – Potassium-doping of the WSe$_2$ surface

In order to experimentally access the single particle spin polarization of the conduction band at the high symmetry K points and the corresponding Σ points between Γ and K, we performed static spin-resolved measurements on a potassium-doped system. The doping process results in an electron transfer from the alkali metal to the WSe$_2$ semiconductor, which increases the chemical potential and ultimately populates the WSe$_2$ conduction band for sufficiently high doping concentrations. We start by discussing the change in the experimentally obtained ARPES spectrum upon potassium doping. First of all, Figure 4 (a) shows the static momentum-resolved photoemission intensity along the Σ–K direction around the Fermi level for pristine WSe$_2$ and for two exemplary steps of potassium doping. The electron doping concentrations have been estimated in accordance with [46]. The photoemission yield is plotted logarithmically to better visualize the band dispersion of the populated conduction band alongside the much more intense valence bands. With successive potassium doping, the valence bands shift to larger binding energies $E_B = -(E - E_F)$ which in Figure 4 (a) is indicated by the black and red dashed lines related to the valence band maximum at the K point. The energetic position of the valence band maximum in respects to the Fermi level $E_F$ for different potassium doping steps is depicted in Figure 4 (b), showing an initial strong shift to larger binding energies, that asymptotically approaches a constant energy position with increasing amount of doping as the surface gets more and more saturated with potassium. The valence band maximum at the K point shifts from $E-E_F = (-1.27 \pm 0.02)$ eV for undoped WSe$_2$ to $E-E_F = (-1.54 \pm 0.02)$ eV for the highest amount of potassium doping, resulting in an overall energy shift of the valence band maximum of $\Delta E = (-0.27 \pm 0.02)$ eV. With consecutive potassium doping, the chemical potential is lifted above the onset of the conduction band at the Σ point, leading to a population of the previously unoccupied conduction band minimum. Besides that, additional dispersive features arise below the conduction band, which are addressed to adsorbate states resulting from the strong potassium doping at the surface. As shown in Figure 4 (c), the size of the indirect K–Σ band gap (measured between the intensity peak maxima) is lowered to $E_g = (1.45 \pm 0.02)$ eV with increasing potassium doping, indicating a stronger shift of the conduction band compared to the valence bands. Such kind of band structure renormalizations upon alkali metal doping are well known and have been investigated before [46-48]. For the highest amount of potassium doping, also the onset of the local conduction band minimum at the K point is weakly populated (see right panel in Figure 4 (a)) with its maximum intensity located at $E - E_F = (-0.12 \pm 0.02)$ eV or $E_K - E_{VBmax} = (1.42 \pm 0.02)$ eV in respects to the valence band maximum. This is particularly notable, as it lies below the energetic position of the conduction band at the Σ point $E_\Sigma - E_{VBmax} = (1.45 \pm 0.02)$ eV, suggesting a shift toward a direct bandgap, as is characteristic of monolayer WSe$_2$ [42]. This shift could result from an intercalation of potassium atoms into the interlayer van der Waals gap between the first and second layer of WSe$_2$, effectively forming a quasi-freestanding monolayer [49, 50]. However, the low intensity of the conduction band at the K point – particularly in comparison to the strong signal at the $\Sigma$ point in Figure 4 (a) – indicates that only the onset of the conduction band is populated, with its actual minimum still located above the Fermi level. Nevertheless, the potassium doping has clearly induced band renormalizations, reducing the energetic separation of the conduction band valleys at the K and $\Sigma$ points, and thereby promoting an overall trend toward a monolayer-like band structure in the bulk material.

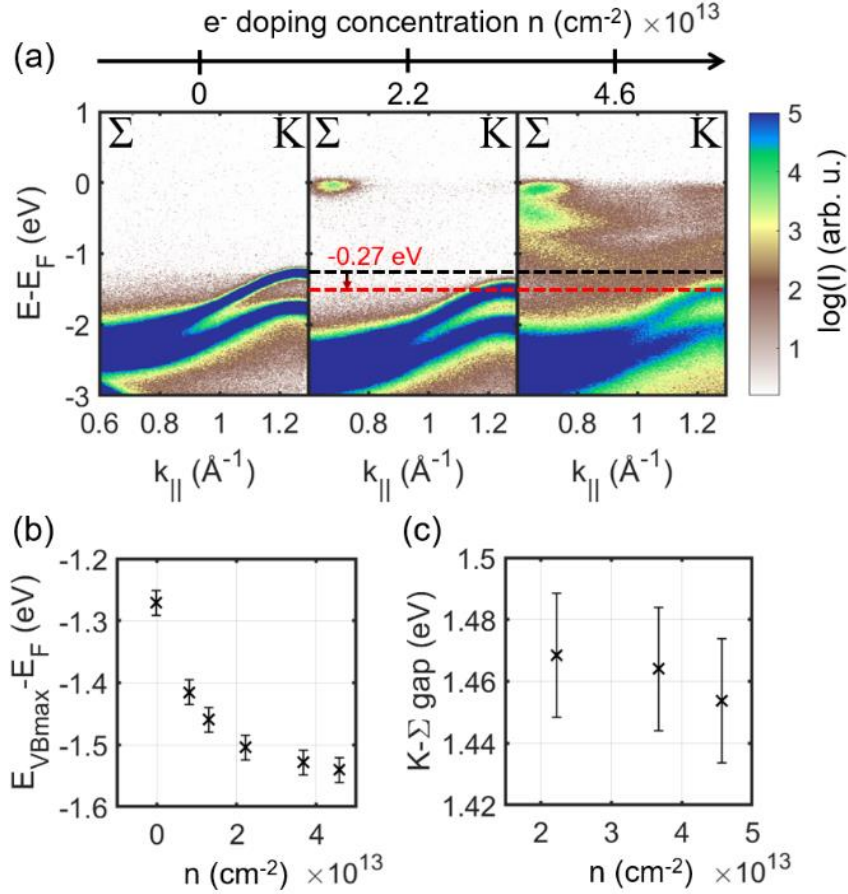

Figure 4: (a) ARPES maps of the band structure of bulk WSe2 along the Σ-K direction for the pristine crystal and two exemplary steps of potassium doping. From the undoped system to the highest amount of potassium doping, the valence band maximum shifts by -0.27 eV, which is indicated by the dashed black and red lines. (b) Shift of the valence band maximum at the K point with increasing potassium doping in respects to the Fermi level. (c) Size of the band gap between the valence band maximum at the K point and the conduction band minimum at the Σ point.

# Transition from Optically Excited to Intrinsic Spin Polarization in WSe$_2$

*Supplementary Information*

S. Hedwig[1], G. Zinke[1], J. Braun[2], B. Arnoldi[1], A. Pulkkinen[3], J. Minár[3], H. Ebert[2], M. Aeschlimann[1,a], B. Stadtmüller[4,b]

[1]Department of Physics and Research Center OPTIMAS, Rheinland-Pfälzische Technische Universität Kaiserslautern Landau, 67663 Kaiserslautern, Germany

[2]Department Chemie, Physical Chemistry, Ludwig-Maximilians-Universität München, 81377 München, Germany

[3]New Technologies Research Centre, University of West Bohemia, 301 00 Plzeň, Czech Republic

[4]Institute of Physics, Experimentalphysik II, Augsburg University, 86159 Augsburg, Germany

[a]Contact author: ma@physik.uni-kl.de
[b]Contact author: benjamin.stadtmueller@uni-a.de

# Spin-Resolved Valence Band and Hole Dynamics in WSe$_2$

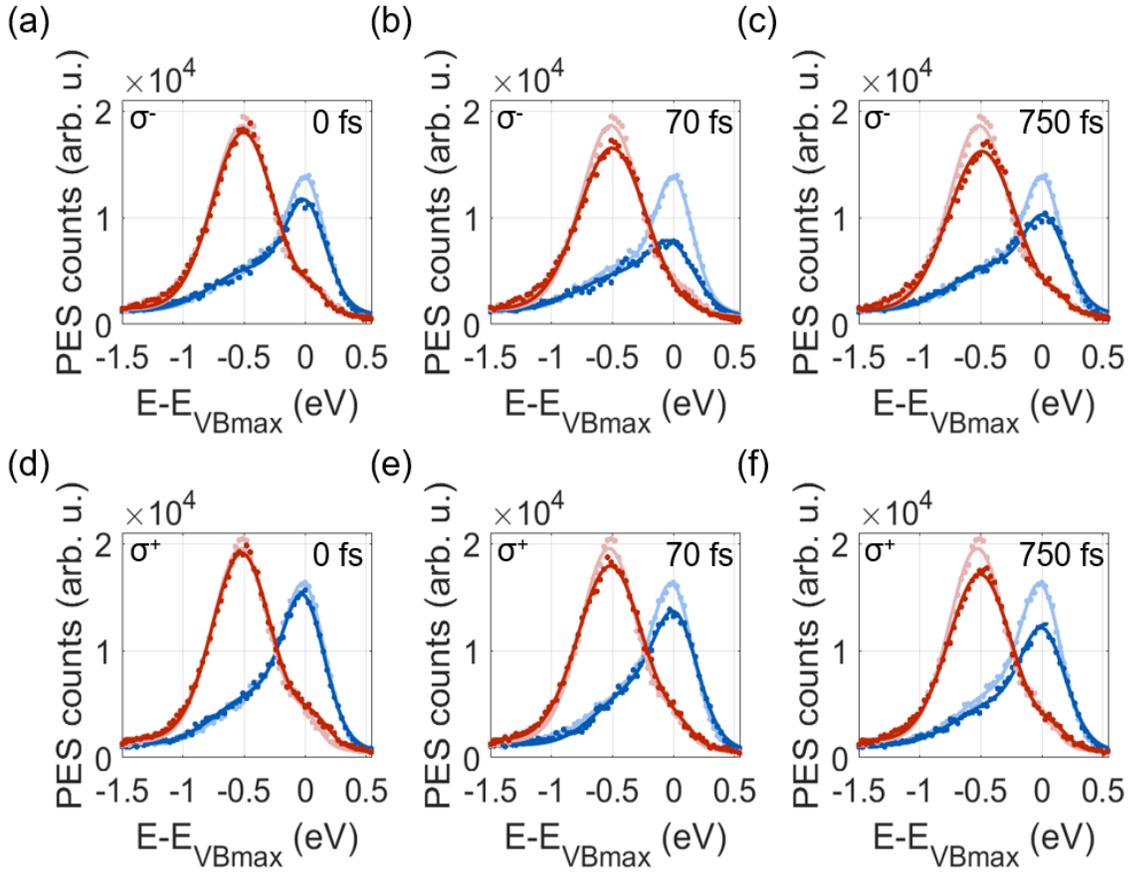

**Figure S1:** Spin-resolved photoemission spectra of the valence bands at the K point for different pump-probe delays after optical excitation with σ⁻-polarized (a) – (c) and σ⁺-polarized light (d) – (f), related to the data shown in Figure 5 (d) and (e) in the main text. The corresponding spectra for each time step (0 fs, 70 fs, 750 fs) are displayed in vivid blue (red) color, representing spin up (down) configuration. The corresponding unexcited spectrum at -500 fs before the optical excitation is depicted in light blue and red color for reference. The dots mark the measured signal, while the solid lines of the same color are fitting curves.

Corresponding to the excited spin carrier populations in the conduction band discussed in the main text, measurements of the valence band, reflecting the hole dynamics, have been carried out and are presented in Figure S1, showing spin-resolved photoemission spectra of the valence bands at the K point for different pump-probe delays after optical excitation with σ⁻-polarized (a) – (c) and σ⁺-polarized light (d) – (f). The spectra are related to the data shown in Figure 3 (d) and (e) in the main text. For each time step (0 fs, 70 fs, 750 fs) the spectra are displayed in vivid blue (red) color, representing spin-up (-down) configuration together with a spectrum at -500 fs before the optical excitation in light blue and red color for reference. The solid lines are fitting curves of the measured data, marked by dots of the same color. For both pump light helicities, a significant drop in intensity at the spin-up polarized valence band maximum can be observed, directly visualizing the electron holes. For $\sigma^-$-polarized excitation in Figures S1 (a) – (c), the drop in intensity is strongest 70 fs after the optical excitation and relaxes back on longer time scales. Thereby, the initial spin-up polarization of ~60% at -500 fs at the valence band maximum is quenched to about ~35% at 70 fs before recovering to ~45% at 750 fs. This is to be expected, since for the chosen light helicity, spin-up carriers of the first layer are excited into the conduction band. For the $\sigma^+$-polarized geometry shown in Figures S1 (d) – (f), however, there is no initial drop in intensity during the optical excitation at 0 fs but rather an evolving decrease that peaks for longer delay times at 750 fs with a spin-up polarization of ~50% compared to ~60% at -500 fs.

Even when the initially excited electron populations at the K points are long gone (see Figure 3 in the main text), holes are still present in the valence band, illustrating the different scattering and spin relaxation dynamics for electrons and holes. Since with $\sigma^+$-polarized light spin-down carriers in the second layer are excited and no spin-up electrons are expected to be directly excited at the observed momentum region, this effect can be ascribed to a combination of a partially not perfectly aligned light polarization, leading to the direct excitation of a small amount of spin-up electrons (spin-down holes) at the observed region, and more importantly, a redistribution of holes via K–K′ exchange spin-flip scattering, leading to a thermalization of the non-equilibrium hole populations. This is best seen by comparing the more equilibrated states in Figures S1 (c) and (f) for the two different excitation geometries, which, despite a different initial excitation (see also Figures 3 (d) and (e) in the main text), thermalize towards the same distribution at 750 fs.

# Scattered Spin Polarization at the Σ and Σ′ Points in WSe$_2$

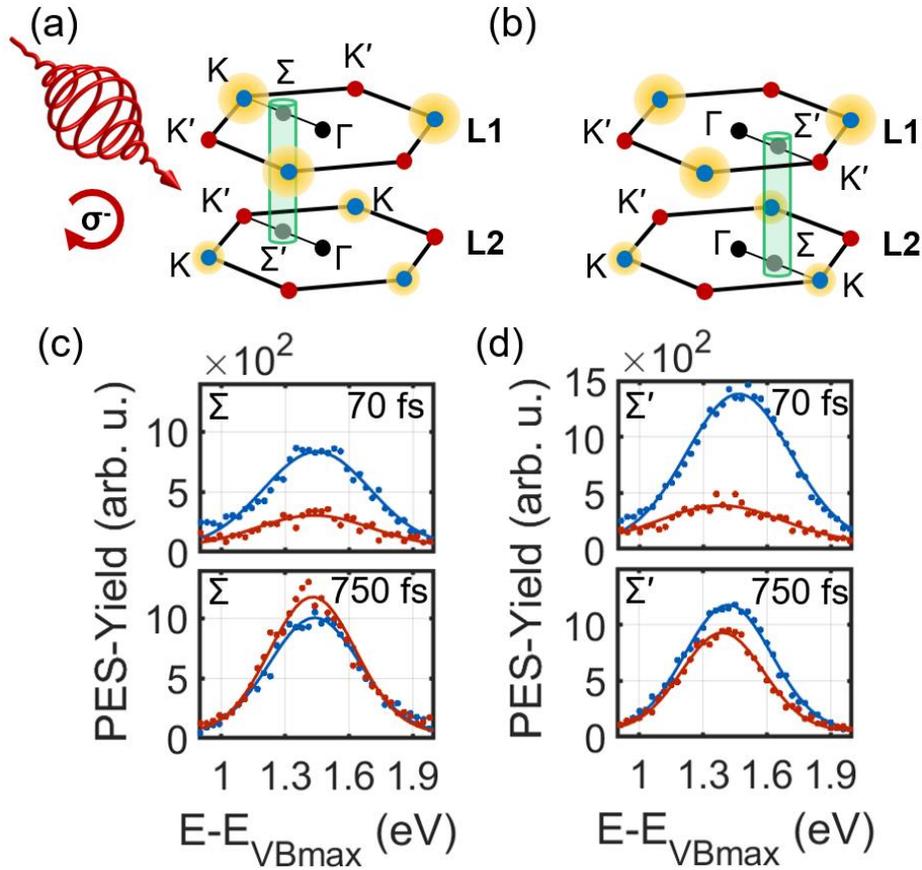

**Figure S2:** (a) – (b) Sketch of the hexagonal surface Brillouin zone of the first (L1) and second (L2) (tri-)layer of WSe$_2$ for excitation with σ$^-$-polarized pump pulses. A yellow corona marks the valleys that are being excited, respectively, while the green area depicts the region in momentum space, where the measurements have been conducted. (c) – (d) Measured spin signal at the Σ point (c), and the Σ′ point (d) for two distinct time steps (70 fs and 750 fs) after optical excitation with σ$^-$-polarized pump pulses at the sketched points in (a) – (b). Blue (red) dots and lines mark spin up (down) data points and their respective Gaussian fit. While at the Σ point (c) the initially scattered spin-up polarization reverses its sign on longer time scales (750 fs) and is negative, no such changes of the spin polarization are observed at the Σ′ point (d), which overall matches the behavior of the intrinsic spin polarization in the potassium doped WSe$_2$ system (see Figure 2 in the main text).

# Thermalization of Excited Carrier Populations

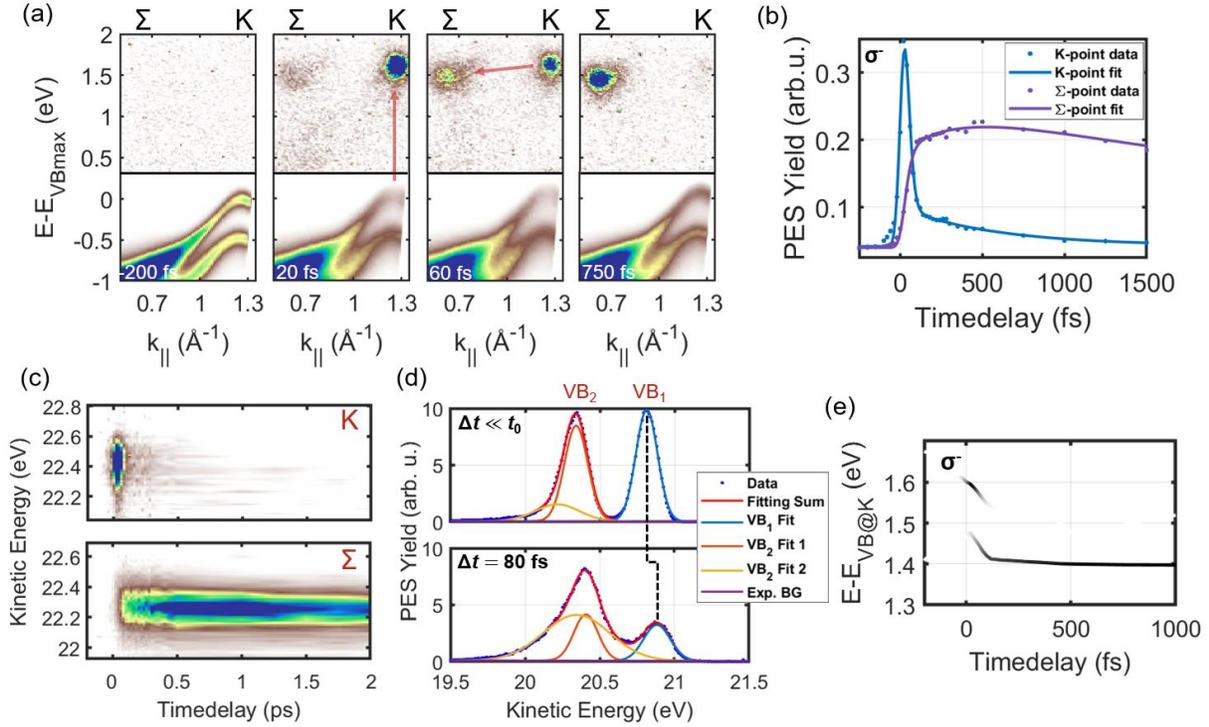

**Figure S3**: (a) ARPES maps for different pump-probe delays (-200 fs, 20 fs, 60 fs and 750 fs) showing an initial excitation of carriers at the K point and a subsequent scattering to the Σ valley. At 20 fs – 60 fs, light coherent replica of the valence bands at $E+h\nu_{pump}$ can be observed. (b) Integrated and fitted intensity signal of the excited populations at the K and Σ point. (c) Energy resolved evolution of the excited populations at the K and Σ valleys of the conduction band. (d) Fitted valence band spectra at two exemplary pump-probe delays before and after the optical excitation. Space charge driven shifts of the whole spectrum are observed upon optical excitation. (e) Shift adjusted energy position of the populations at the K (upper signal) and Σ point (lower signal) obtained from Gaussian fitting of (c). The energetic position is referenced to the time dependent energy position of the valence band maximum in (d). The grayscale of the solid line reflects its relative intensity value ranging from white (no intensity) to black (maximum intensity). A quick thermalization of both the excited population at the K point and the scattered population at the Σ valley within ~100 fs can be observed.

Figure S3 (a) shows a typical tr-ARPES experiment performed on bulk $WSe_2$ in the Σ–K region using $\sigma^-$-polarized pump pulses at four different pump-probe time delays, -200 fs before the excitation and 20 fs, 60 fs and 750 fs after the optical excitation. The contrast above the black horizontal line has been adjusted to better visualize the excited populations alongside the valence bands. Shortly after the optical excitation at 20 fs, a drop of intensity at the valence band maximum can be seen, visualizing the holes via the absence of photoelectrons in that region. At the same time a population arises in the former unoccupied conduction band at the K point, which has been excited from the valence band, indicated by a red arrow. A very weak population can already be seen at the Σ point of the conduction band, that has been scattered from the K point. Additionally, a weak replication of the spin-split upper valence bands is visible. These replicas occur above the initial valence band position, shifted exactly by the pump photon energy $h\nu_{pump} = 1.59$ eV and are only observable during the presence of the pump pulse. Therefore, these features can be addressed to photon-dressed states coherently coupling to the pump light field [22, 35-39]. At 60 fs, a larger fraction of the excited carriers has scattered to the Σ point, while the hole population in the valence band at the K point still persists. Even at longer times, 750 fs after the optical excitation, there are still holes observable in the valence band and almost all carriers initially excited at the K point have undergone intervalley scattering to the Σ point or recombination with holes in the valence band. These results are qualitatively consistent with several previous experiments performed on WSe$_2$ and other TMDs [14, 22, 35, 40]. The evolution and intervalley scattering of the excited carrier population from the K to the Σ point can best be seen looking at the momentum and energy integrated signal in the region around the K and Σ point each, which is shown in Figure S3 (b) for an excitation with σ⁻-polarized light. Figure S3 (c) shows the evolution of the momentum-integrated kinetic energy distribution of photoelectrons at the K and Σ point, excited with σ⁻-polarized light. A rapid thermalization of the hot carrier populations towards lower energies via electron-electron and electron-phonon intraband scattering [22] can be seen. In order to quantify this thermalization towards the conduction band minima, energy shifts of the entire band structure due to time-zero space charge effects [172, 173] have to be taken into account. These shifts occur equally for all bands and last longer than the 10 ps time frame examined. Therefore, the energetic postitions of the populations at the K and Σ point are referenced to the upper valence band maximum VB1 for every pump-probe delay step. A cut through the upper valence bands at the K point is depicted in Figure S3 (d) showing a drop of intensity at VB1 80 fs after the optical excitation, as electrons are excited into the conduction band, leaving behind a hole in the valence band. Besides that, an energy shift of $\Delta E = (70 \pm 20)$ meV compared to the unexcited spectrum can be observed. For each time delay step, the spectrum was fitted using Gaussian peak functions in combination with an exponential background fit. Two Gaussian curves had to be used in order to model the slightly asymmetric shape of VB2. The time-dependent peak positions of VB1 can then be used to relate the energetic positions of the excited carrier populations and visualize their evolution without the influence of space charge shifts superimposing the carrier relaxation. The result is displayed in Figure S3 (e), where the centers of the energy distributions in (c) were determined by Gaussian peak fitting for each individual time step and related to the respective energetic position of the maximum of VB1. Additionally, the overall intensity of the populations is reflected by the grayscale of the solid line, ranging from white (no intensity) to black (maximum intensitiy). The upper signal setting in at around $E-E_{VB@K} = (1.61 \pm 0.02)$ eV refers to the hot carrier population at the K point which thermalizes to $E-E_{VB@K} = (1.53 \pm 0.02)$ eV within 140 fs upon optical excitation before becoming indistinguishably weak due to recombination with holes in the valence band and intervalley scattering to the Σ point. This value can be assumed to be the bottom of the conduction band valley at the K point. The lower lying signal of the scattered carriers at the Σ point starts at $E-E_{VB@K} = (1.48 \pm 0.02)$ eV and is quickly redistributed to $E-E_{VB@K} = (1.41 \pm 0.02)$ eV, where it stays stable for longer than 10 ps, the experimentally observed time frame, only losing minimal intensity. The obtained value of $E-E_{VB@K} = (1.41 \pm 0.02)$ eV for the energetic position of the global conduction band minimum at the Σ point lies within the boundaries of the statically measured $E_{\Sigma}-E_{VBmax} = (1.45 \pm 0.02)$ eV for the K–Σ gap in potassium-doped $WSe_2$ presented in Figure 4 in the End Matter of the main text.

# References SI